\documentclass[epj]{svjour}
\usepackage{graphics}
\usepackage[latin1]{inputenc}
\begin{document}
\title{Superconductivity at $T_c$ = 44 K in Li$_x$Fe$_2$Se$_2$(NH$_3$)$_y$}
\author{E.-W. Scheidt\inst{1}, V. R. Hathwar\inst{1}, D. Schmitz\inst{1}, A. Dunbar\inst{1}, W. Scherer\inst{1}, V. Tsurkan\inst{2,} \inst{3}, J. Deisenhofer\inst{2}, A. Loidl\inst{2}
}                     
%
%
\institute{CPM, Institute of Physics, University of Augsburg, 86157
Augsburg, Germany \and Center for Electronic Correlations and
Magnetism, Institute of Physics, University of Augsburg, D 86159,
Augsburg, Germany \and Institute of Applied Physics, Academy of
Sciences of Moldova, MD 2028, Chisinau, R. Moldova}

\date{Received: date / Revised version: date}
%
\abstract{Following a recent proposal by Burrard-Lucas et al.
[unpublished, arXiv: 1203.5046] we intercalated FeSe by Li in liquid
ammonia. We report on the synthesis of new
Li$_x$Fe$_2$Se$_2$(NH$_3$)$_y$ phases as well as on their magnetic
and superconducting properties. We suggest that the superconducting
properties of these new hybride materials appear not to be
influenced by the presence of electronically-innocent Li(NH$_2$)
salt moieties. Indeed, high onset temperatures of 44 K and shielding
fractions of almost 80\% were only obtained in samples containing
exclusively Li$_x$(NH$_3$)$_y$ moieties acting simultaneously as
electron donors and spacer units. The $c$-axis of the new
intercalated phases is strongly enhanced when compared to the
alkali-metal intercalated iron selenides $A_{1-x}$Fe$_{2-y}$Se$_2$
with $A$ = K, Rb, Cs, Tl with $T_c$ = 32~K.
} 
\maketitle
\section{Introduction}
\label{intro} The discovery of iron-based superconductors in 2008
\cite{Kamihara08} had boosted hopes to find compounds that would
rival the $T_c$ records of the copper-based superconductors
\cite{Basov11}. The highest transition temperatures of Fe based
superconductors to date are in the vicinity of 56~K and were
reported already within the first year after the initial discovery
\cite{CaoWang08,Cheng09}. Meanwhile a variety of families has been
identified \cite{Rotter08,Kant10,Wang08,Ogino09} all of which share
a common structural unit, namely Fe$_2$As$_2$ or Fe$_2$Se$_2$ layers
which are responsible for carrying superconductivity (see
\cite{Johnston2010} and \cite{Stewart2011} for recent reviews). When
the binary chalcogenide superconductor FeSe with a critical
temperature of about 8~K appeared on stage \cite{Hsu08}, not only a
simple model systems to study the origin and mechanism of
superconductivity was found, but also the expectation had been
refueled to raise $T_c$ above the temperature of liquid nitrogen:
First, substitution of Se by Te increased $T_c$ to about 15~K
\cite{Fang08,Yeh08,Tsurkan11B}. Then the critical temperature of
FeSe was found to increase to 37~K under pressure \cite{Medvedev09}.
A further milestone in the evolution of Fe-based superconductors was
set by the synthesis of $A_{1-x}$Fe$_{2-y}$Se$_2$ single crystals
with $A$ = K, Rb, Cs, Tl and $T_c\sim$ 32~K \cite{Guo10,Ye11}. In
addition to being superconducting these compounds exhibit
iron-vacancy order in a $\sqrt{5}$*$\sqrt{5}$*1 supercell below 580
K and local moment antiferromagnetism with large ordered moments and
magnetic ordering temperatures between 470~K - 560~K
\cite{Ye11,Bao11}. Recently, ample experimental evidence showed that
these systems, which are sometimes referred to as the 245-family,
are phase separated
\cite{Ricci11,Li12,Ksenofontov11,Texier12,Charnukha12} where thin
metallic Fe$_2$Se$_2$ sheets with no Fe vacancies \cite{Texier12}
carry superconductivity at low temperatures and alternate with
insulating antiferromagnetic layers \cite{Charnukha12}. The fact
that these natural heterostructures behave like ideal single
crystals with metallic sheets almost epitaxially intergrown within
the insulating antiferromagnet is best documented by the observation
of the characteristic spin excitation modes of unconventional
superconductors at well defined Q-values in reciprocal space by
inelastic neutron scattering \cite{Park11,Friemel12}. Applying
pressure leads to a suppression of both antiferromagnetism and
superconductivity in these systems at the same critical pressure of
approximately 6 GPa \cite{Gooch11,Guo12,Ksenofontov12}. In addition,
a new superconducting phase was found to emerge at pressures beyond
12 GPa with $T_c$ = 48 K \cite{Sun12}.

Recently, reports of spurious superconductivity around 40 K appeared
\cite{Fang11,Wang11,Zhang12}, and it was claimed that $T_c$ is
increasing with increasing distance of the FeSe layers to a critical
temperature of 44 K \cite{Zhang12}. A completely new route for the
Fe chalcogenide superconductors was put forward by Ying et al.
\cite{Ying12} and Burrard-Lucas et al. \cite{Burrard12} who
pioneered the intercalation of FeSe by earth-alkali and rare earth
elements in liquid ammonia leading to $T_c$ = 43~K.

Following a modified intercalation approach we obtained
superconducting samples with a critical onset temperature of
superconductivity of 44~K and a diamagnetic shielding signal
corresponding to about 80\% of the sample volume. The study by
Burrand-Lucas et al. stresses the role of intercalated lithium ions,
lithium amide Li(NH$_2$) and ammonia as spacer layers. In the
following we will show, however, that (i) the presence of the
lithium amide as electronically-innocent guest species  is no
prerequisite for the onset of superconductivity while (ii) the
superconducting properties are controlled by electronic doping and
lattice expansion due to the presence of the Li$_x$(NH$_3$)$_y$
moieties.

These results could be the starting point to employ tailor-made
electronic donor molecules (e.g. metallocenes) which allow for a
systematic variation of the donor capabilities of the guest species
and the interlayer separation in Fe$_2$Se$_2$ hybrids and thus a
systematic control of the critical temperatures.



\begin{figure}[b]
\centering
\resizebox{1.0\columnwidth}{!}{%
  \includegraphics{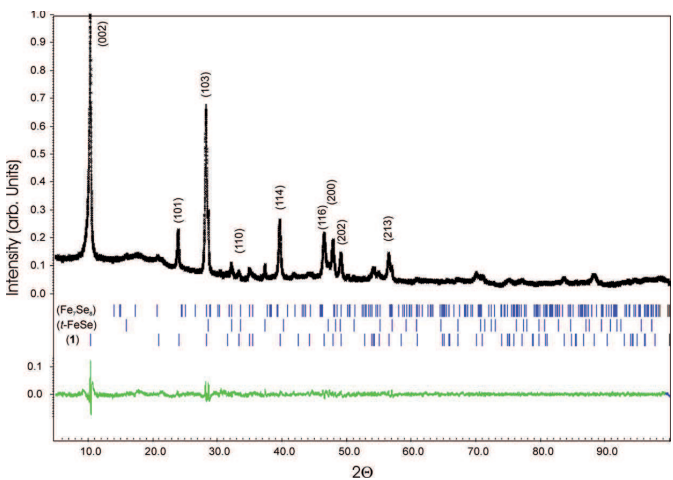}
}
\caption{Powder diffraction pattern of
Li$_{1.8}$Fe$_2$Se$_2$(NH$_3$)[Li(NH$_2$)]$_{0.5}$ (\textbf{1}) at room temperature.
The intensity was normalized to the intensity of the (002) reflection. The
result of a Le-Bail fit is indicated as a solid line. The resulting
difference spectrum is indicated at the bottom of the figure. The
calculated and allowed Bragg reflections of the parent compound
(tetragonal Fe$_2$Se$_2$), Fe$_7$Se$_8$ impurities (traces) and the
intercalated hybride material (\textbf{1}) are indicated by vertical bars.}
\label{fig:xrd}       
\end{figure}

\begin{figure*}[t]
\centering \resizebox{1.8\columnwidth}{!}{
  \includegraphics{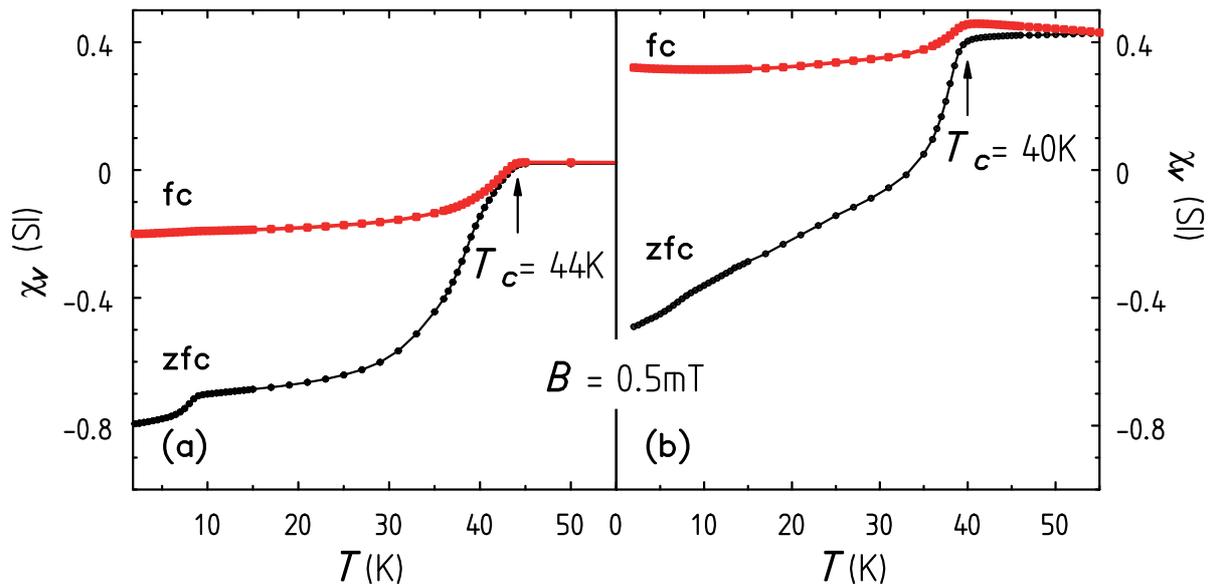}
}
\caption{Temperature dependence of the magnetic susceptibility of
intercalated FeSe samples as obtained in zero-field-cooled and
field-cooled runs. In field cooling cycles and as probing dc fields,
external magnetic fields of 0.5 mT have been used. (a) Volume
susceptibility of the batch with the lowest normal-state
susceptibility Li$_{0.5}$Fe$_2$Se$_2$(NH$_3$)$_{0.6}$. (b) Volume
susceptibility of the batch with the highest normal-state
susceptibility Li$_{1.8}$Fe$_2$Se$_2$(NH$_3$)[Li(NH$_2$)]$_{0.5}$.
The superconducting onset temperatures are indicated by arrows.}
\label{fig:susc_dia}       
\end{figure*}

\section{Experimental details}

\begin{figure*}[t]
\centering \resizebox{1.8\columnwidth}{!}{
\includegraphics{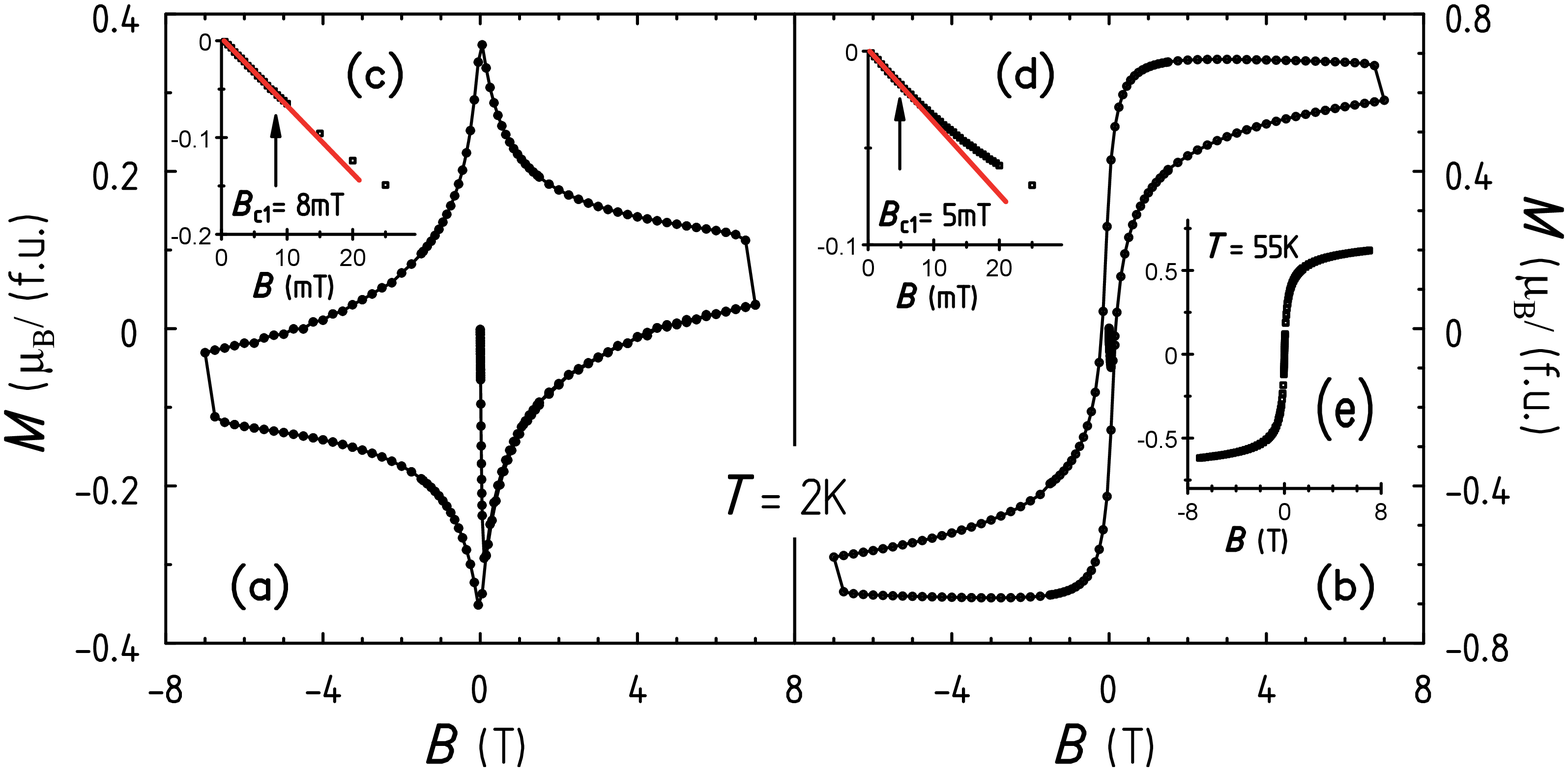}
}

\caption{Magnetic hysteresis loop of intercalated FeSe samples at 2
K: (a) of the batch with the lowest paramagnetic background
(Li$_{0.5}$Fe$_2$Se$_2$(NH$_3$)$_{0.6}$) and (b) of the batch with
the highest paramagnetic background
(Li$_{1.8}$Fe$_2$Se$_2$(NH$_3$)[Li(NH$_2$)]$_{0.5}$). The insets (c)
and (d) display an expanded region of the magnetization curves which
allows to estimate the lower critical field, $H_{c1}$(2K) = 8mT and
5mT  in Li$_{0.5}$Fe$_2$Se$_2$(NH$_3$)$_{0.6}$ and
Li$_{1.8}$Fe$_2$Se$_2$(NH$_3$)[Li(NH$_2$)]$_{0.5}$, respectively.
Inset (e) exhibits the hysteresis loop at 55 K reflecting the
ferrimagnetic contribution of the samples with the highest magnetic
background.} \label{fig:mag}
\end{figure*}

\label{sec:1} Polycrystalline samples of tetragonal FeSe were
synthesized from high purity Fe pieces (99,99\%) and Se (99,999\%)
shots. Stoichiometric mixtures of the starting materials were placed
in double-wall ampoules and were slowly heated to 1100°C, kept at
this temperature for 48 h and then cooled with a rate of 60°C/h to
410°C. At 410°C the ampoules were kept for 100 h and then quenched
in ice water. X-ray diffraction and SQUID measurements documented
single phase character of the materials lacking any impurity phases
and displaying the appropriate tetragonal space group $P4/nmm$, with
lattice constants $a = b = 0.3771$~nm and $c = 0.5524$~nm.
Furthermore,  a well-defined transition into the superconducting
state with an onset temperature of 9.5~K was identified. The high
transition temperature and the ratio of $c/a = 1.4648$ signal a
composition close to Fe$_{1.02}$Se \cite{Queen09}. We note, that
careful inspection of powderized samples of tetragonal FeSe revealed
its metastable character. Indeed, a tribochemical transition and
formation of Fe$_7$Se$_8$ as a ferrimagnetic impurity
\cite{Adachi61} is observed upon grinding of tetragonal FeSe
samples. Hence, excessive grinding of the samples should be avoided
before the subsequent intercalation reaction in liquid ammonia.

The various intercalation reactions of tetragonal FeSe host lattices
with lithium were carried out under inert gas conditions in liquid
ammonia employing a Schlenk line.  In order to prevent the formation
and intercalation of significant amounts of lithium amide rather
small batches of FeSe (100-600 mg) were intercalated under avoidance
of high lithium concentrations in liquid ammonia. Hence, an excess
of Li metal (99,9\%; Sigma Aldrich) was avoided to synthesize
amide-free Li$_x$Fe$_2$Se$_2$(NH$_3$)$_y$ hybride materials.
Furthermore, the cooling bath temperature was always kept at about
-75°~C during the intercalation period (typically 1-4 hours) and the
removal of the remaining NH$_3$ solvent was accomplished via
condensation into a cooling trap using a vacuum pump.  The dry
sample was allowed to warm up before the transferral to a glove box
(argon inert gas) which is equipped with an inlet system adopted for
the sample holders of the subsequent magnetic measurements.  Hence,
all sample manipulations during synthesis and physical property
measurements were strictly performed under inert gas condition. Two
samples with different amounts of lithium were synthesized by this
approach: Li$_{0.5}$Fe$_2$Se$_2$(NH$_3$)$_{0.6}$ and
Li$_{0.9}$Fe$_2$Se$_2$(NH$_3$)$_{0.5}$. Elemental analysis of the
two samples yielded N:H ratios of 1:3.06 and 1:2.97, respectively,
in line with the successful intercalation of Li(NH$_3$) moieties and
avoidance of any significant Li(NH$_2$) impurity phases. Both
samples yielded $T_c$-values of 44~K and shielding fractions as high
as 80\%. This result clearly suggests that the Li(NH$_2$) impurities
found in the materials prepared by Burrard-Lucas et al. do not
trigger the superconducting properties of the LiFe$_2$Se$_2$ hybride
phases.

Indeed, a control experiment using an excess of lithium during the
intercalation process yielded a product containing significant
amounts of Li(NH$_2$) with the formal stoichiometry of
Li$_{1.8}$Fe$_2$Se$_2$(NH$_3$)[Li(NH$_2$)]$_{0.5}$ but yielding a
lower $T_c$ of 40~K. Since the Fe$_2$Se$_2$ parent lattice only
provides voids to accommodate formally one  NH$_3$/NH$_{2}^-$ moiety
per formula unit it remains to be seen, whether Li(NH$_2$) represent
a true guest species or just an impurity phase. We therefore suggest
that the electronically inert Li(NH$_2$) moiety will not contribute
to the physical properties of these hybride materials.

Magnetization measurements were performed in a magnetic property
measurement system MPMS-7 (Quantum Design), in a temperature range
between 2~K to 300~K and in magnetic fields up to 7~T. All powder
samples were mounted in a special Kel-F sample holder, which has a
cylindrical hole with a diameter of 3~mm and a height of 3~mm. The
samples were prepared in argon atmosphere and transferred to the
magnetometer via an argon-lock. To determine the volume
susceptibility of the intercalated samples, the density $\rho$ was
derived from the unit cell volume based on the powder diffraction
and the analytical data. The calculated density for
Li$_{0.9}$Fe$_2$Se$_2$(NH$_3$)$_{0.5}$ is 3.937\,g/cm$^3$ and for
Li$_{1.8}$Fe$_2$Se$_2$(NH$_3$)[Li(NH$_2$)]$_{0.5}$ the value is
4.24\,g/cm$^3$.

\section{Experimental results and discussion}
\label{sec:2}

Phase identification and purity of the parent lattices and of the
intercalated hybrid materials was controlled by powder diffraction
studies using a Image Plate Guinier Camera G670 (Huber) and
monchromatized CuK$_{\alpha 1}$ radiation with $\lambda =
1.540598$~Å. A flat sample holder was employed and the inherent air
and moisture sensitive samples were prepared inside a glove box. The
samples were sealed in between two Mylar foils to prevent sample
decomposition. Phase analysis and lattice parameter refinements were
performed using the Le-Bail method \cite{LeBail04}.

In Fig.~\ref{fig:xrd} the x-ray diffraction pattern is shown for
Li$_{1.8}$Fe$_2$Se$_2$(NH$_3$)[Li(NH$_2$)]$_{0.5}$. All Bragg
intensities of the intercalated species could be indexed by a body
centered tetragonal cell with \textit{I4/mmm} symmetry in agreement
with the study by Burrard-Lucas et al. \cite{Burrard12}. The lattice
parameters were found to be rather insensitive to the amount and
ratio of intercalated Li(NH$_3$)/Li(NH$_2$) fractions, for example,
$a$ = $b$ = 0.38273(6)\,nm and $c$ = 1.6518(3)\,nm (space group
\textit{I4/mmm}) for Li$_{0.9}$Fe$_2$Se$_2$(NH$_3$)$_{0.5}$ and $a$
= $b$ = 0.379607(8)\,nm and $c$ = 1.69980(11)\,nm in the case of
Li$_{1.8}$Fe$_2$Se$_2$(NH$_3$)[Li(NH$_2$)]$_{0.5}$. Compared to the
FeSe starting material the in-plane lattice constants are almost the
same, being expanded by less than 0.3\%. The $c$-axis however is
enlarged by more than a factor of 3. Compared to the 245 compounds
\cite{Ye11} the in-plane lattice constants ($a\sqrt{5}$) are
slightly smaller but the $c$-axis is dramatically increased, a fact
which strongly points toward the importance of the FeSe layer
separation along $c$ to enhance the $T_c$ values. However, one has
to keep in mind that the alkali intercalated 245 compounds exhibit
critical temperatures of approximately 32\,K, independent of the
magnitude of the $c$-axis lattice change, which increases from
1.4\,nm in the potassium containing compounds to 1.53\,nm in the Cs
intercalated compounds \cite{Ye11}.

The temperature dependent magnetic susceptibilities of two
intercalated Fe$_2$Se$_2$ samples are depicted in
Fig.~\ref{fig:susc_dia} showing the respective zero-field-cooled
(zfc) and field-cooled (fc) runs. The earth magnetic field was
compensated during the zfc sequences (down to 2\,K) by a procedure
described in detail in Ref.~\cite{Presnitz12}. In the subsequent
heating run we applied a small magnetic field ($B$ = 0.5\,mT) to
record the magnetization data up to 55\,K. This procedure reflects
the complete shielding effect of the sample at low temperature
($\chi_V$ = -1 in the ideal case), whereas the fc measurements
account for the Meissner expulsion. Figure~\ref{fig:susc_dia} shows
the temperature dependence of the volume susceptibility $\chi_V(T)$
 of two selected samples, namely Li$_{0.5}$Fe$_2$Se$_2$(NH$_3$)$_{0.6}$ with the lowest
normal-state susceptibility (Fig.~\ref{fig:susc_dia}a) and
Li$_{1.8}$Fe$_2$Se$_2$(NH$_3$)[Li(NH$_2$)]$_{0.5}$ with the largest
normal-state susceptibility values (Fig.~\ref{fig:susc_dia}b).

In case of Li$_{0.5}$Fe$_2$Se$_2$(NH$_3$)$_{0.6}$
(Fig.~\ref{fig:susc_dia}a) we find a shielding fraction of about
80\% as observed in the zfc measurements and observe the $\chi_V(T)$
signature of another superconducting transition below 10~K which we
ascribe to traces of the non-intercalated parent compound FeSe. The
fc experiments point towards a small lower critical field and a
moderate pinning effect leading to a Meissner phase which amounts
approximately 20\% of the sample volume. Both volume fractions seem
to be significant and rather large when compared to the results by
Ying et al. \cite{Ying12} and Burrard-Lucas et
al.\,\cite{Burrard12}. A well-defined onset of superconductivity in
Li$_{0.5}$Fe$_2$Se$_2$(NH$_3$)$_{0.6}$ appears close to 44\,K in
both, the zfc and fc experiments. This is one of the highest
transition temperatures reported so far in the iron-selenides at
ambient pressure. For temperatures $T > T_c$ we find a small and
almost vanishing paramagnetic Pauli-like magnetic susceptibility
only.

In case of Li$_{1.8}$Fe$_2$Se$_2$(NH$_3$)[Li(NH$_2$)]$_{0.5}$
(Fig.~\ref{fig:susc_dia}b) the relative behavior of the fc and zfc
susceptibility values is similar to the amid free sample
(Fig.~\ref{fig:susc_dia}a) with exception of the somewhat lower
critical temperature $T_c$ = 40~K and the large normal-state
susceptibility contribution of $\chi_V$ = 0.4. Accordingly, the fc
curve is completely shifted to positive susceptibility values.
Subtracting this normal-state susceptibility of 0.4, we find that
both samples exhibit shielding fractions of about 80\% as observed
in the zfc measurements.

We will outline below that this large normal-state susceptibility
might originate from a ferrimagnetic impurity (Fe$_7$Se$_8$) which
is absent in Li$_{0.5}$Fe$_2$Se$_2$(NH$_3$)$_{0.6}$ but present in
Li$_{1.8}$Fe$_2$Se$_2$(NH$_3$)[Li(NH$_2$)]$_{0.5}$ as revealed by
the diffraction pattern (Fig.~\ref{fig:xrd}. A similar
susceptibility contribution has been observed in the Li(ND$_2$)
containing sample
Li$_{0.6}$Fe$_2$Se$_2$(ND$_2$)$_{0.2}$(ND$_3$)$)_{0.8}$ reported by
Burrad-Lucas et al. \cite{Burrard12}, which displays the highest
superconducting volume fraction (40-50\%) in their earlier report.

In the latter case the authors found 10.3\% hexagonal FeSe
impurities via Rietveld analysis \cite{Burrard12}. Consequently, the
reduction or complete avoidance of FeSe impurities during sample
preparation might provide one of the key control parameter of the
superconducting properties of the intercalated FeSe species.
Accordingly, in all compounds there seems to be a correlation
between the superconducting transition temperature and the positive
normal-state susceptibility values.



In order to elucidate the relationship between the superconducting
state and the high positive normal-state susceptibility,
magnetization measurements at 2 K and 55K were performed. The
magnetization versus magnetic field is shown in Fig.~\ref{fig:mag}a
for Li$_{0.5}$Fe$_2$Se$_2$(NH$_3$)$_{0.6}$ and in
Fig.~\ref{fig:mag}b for the system
Li$_{1.8}$Fe$_2$Se$_2$(NH$_3$)[Li(NH$_2$)]$_{0.5}$ in fields up to
7~T. In case of the LiNH$_2$ free sample (Fig.~\ref{fig:mag}a) the
intercalated compound exhibits the typical hysteresis loop of a
type-II superconductor. As in most of the 245 iron selenides and
probably as a fingerprint of 2D superconductors the lower critical
field is close to zero and hence, in these systems the Meissner
phase only exists close to zero external fields. The small asymmetry
of the magnetization hints for a a small magnetic contribution.

This contribution is in case of the LiNH$_2$ containing sample
(Fig.~\ref{fig:mag}b) clearly identified as a ferrimagnetic
impurity. The hysteretic loop at 55 K (Fig.~\ref{fig:mag}e) shows
this underlying magnetic contribution, which is most likely due to
the presence of the Fe$_7$Se$_8$ impurities and/or additional free
Fe-ions. This phase is ferrimagnetic with a saturation magnetism of
0.2 $\mu_B$/Fe-atom and a critical temperature $T_K$ = 425 K
\cite{Adachi61}. From the linear slope of this magnetization curve
between -10 and 10~mT a susceptibility contribution can be derived
which is in good agreement with the observed magnetic normal-state
contribution in Fig.~\ref{fig:susc_dia}(b). In order to estimate the
lower critical field $B_{c1}$ of the samples linear fits to the
initial slopes (solid lines) were performed as depicted in the
insets of Figs.~\ref{fig:mag}(c) and (d). $B_{c1}$ at 2K is
determined by the deviation of the magnetization data from this
straight line resulting in $B_{c1}$(2K) = 0.8 $\pm$ 0.1 mT for
Li$_{0.5}$Fe$_2$Se$_2$(NH$_3$)$_{0.6}$ and 0.5 $\pm$ 0.1 mT in the
case of Li$_{1.8}$Fe$_2$Se$_2$(NH$_3$)[Li(NH$_2$)]$_{0.5}$.

In summary, we synthesized superconducting hybride materials
Li$_x$Fe$_2$Se$_2$(NH$_3$)$_y$ via intercalation of lithium in
liquid ammonia. Li$_{0.5}$Fe$_2$Se$_2$(NH$_3$)$_{0.6}$ with a
maximal superconducting onset temperature of 44~K is almost free of
magnetic impurities with a normal-state susceptibility close to
zero. The Meissner fraction of this compound is about 20\% and the
shielding fraction close to 80\%. The enhancement of the critical
temperature results from the significant increase of the $c$-axis
lattice parameter and electron doping via lithium ions. In addition,
we synthesized Li$_{1.8}$Fe$_2$Se$_2$(NH$_3$)[Li(NH$_2$)]$_{0.5}$.
Here we found a reduced superconducting transition temperature and a
significant amount of magnetic impurities. We hope that these
results are the starting point to systematically vary the separation
of the Fe$_2$Se$_2$ layers by introducing tailored electronic donor
molecules.

This work has partly been supported by the DFG via the SPP 1458 (DE
1762/1-1), by TRR 80 (Augsburg-Munich) and the SPP 1178 (SCHE
487/8-3).

\end{document}